\documentclass[twocolumn]{aastex63}
\usepackage{hyperref}
\usepackage{courier}
\usepackage[T1]{fontenc}
\usepackage{gensymb}
\usepackage{amssymb}
\usepackage{amsmath}
\usepackage{url}

\usepackage{soul}

\usepackage{textcomp,gensymb}

\shorttitle{3FHL Identification}
\shortauthors{Silver et al.}

\begin{document}

\title{Identifying the 3FHL Catalog. IV. \textit{Swift} Observations of Unassociated \textit{Fermi}-LAT 3FHL Sources}
\author{R. Silver}
\affiliation{Department of Physics and Astronomy, Clemson University,  Kinard Lab of Physics, Clemson, SC 29634, USA}

\author{S. Marchesi}
\affiliation{Department of Physics and Astronomy, Clemson University,  Kinard Lab of Physics, Clemson, SC 29634, USA}
\affiliation{INAF $-$ Osservatorio di Astrofisica e Scienza dello Spazio di Bologna, Via Piero Gobetti, 93/3, 40129, Bologna, Italy}

\author{L. Marcotulli}
\affiliation{Department of Physics and Astronomy, Clemson University,  Kinard Lab of Physics, Clemson, SC 29634, USA}

\author{A. Kaur}
\affiliation{Department of Astronomy and Astrophysics, 525 Davey Lab, Pennsylvania State University, University Park, PA 16802, USA}

\author{M. Rajagopal}
\affiliation{Department of Physics and Astronomy, Clemson University,  Kinard Lab of Physics, Clemson, SC 29634, USA}

\author{M. Ajello}
\affiliation{Department of Physics and Astronomy, Clemson University,  Kinard Lab of Physics, Clemson, SC 29634, USA}

\begin{abstract}
    The \textit{Fermi} Large Area Telescope (\textit{Fermi}-LAT) 3FHL catalog is the latest catalog of $ > 10$ GeV sources and will remain an important resource for the high-energy community for the foreseeable future. Therefore, it is crucial that this catalog is made complete by providing associations for most sources. In this paper, we present the results of the X-ray analysis of 38 3FHL sources. We found a single bright X-ray source in 20 fields, two sources each in two fields and none for the remaining 16. The analysis of the properties of the 22 3FHL fields with X-ray sources led us to believe that most ($\sim19/22$) are of extra-galactic origin. A machine-learning algorithm was used to determine the source type and we find that 15 potential blazars are likely BL Lacertae objects (BL Lacs). This is consistent with the fact that BL Lacs are by far the most numerous population detected above $ > 10$ GeV in the 3FHL. \\
\end{abstract}

\section{Introduction} 

\indent \indent 
Gamma rays can provide insight into the most powerful objects in the universe. The very first sensitive census of the $\gamma$-ray sky came in 1993 when the Energy Gamma Ray Experiment Telescope \citep[EGRET;][]{Fichtel} on the Compton Gamma Ray Observatory (CGRO) completed a survey of the $\gamma$-ray sky above 50 MeV. This revealed many high-energy astrophysical objects, such as active galactic nuclei (AGN), gamma-ray bursts, supernova remnants, and pulsars. In 2008, the Large Area Telescope \citep[LAT;][]{Atwood09} on board the \textit{Fermi} gamma-ray space telescope took the next step in $\gamma$-ray astrophysics with its improved sensitivity and resolution over EGRET by factors of 100 and 3, respectively. \textit{Fermi}-LAT is sensitive to the detection of hard-spectrum sources (emission $> 10$ GeV) as demonstrated in the 1FHL (514 objects detected above 10 GeV) and the 2FHL (360 objects detected above 50 GeV).
The newest catalog in this series, the 3FHL \citep{Ajello17}, provided a significant improvement with the detection of 1556 sources between 10 GeV and 2 TeV relying on the first 7 years of LAT data. \\ \indent
However, upon release, the 3FHL had 200 sources listed as either unknown (i.e., associated with a source of unknown nature) or lacking a firm association in any other wavelength. The median positional resolution of 2.3$\arcmin$ hinders the easy identification of the counterpart. Finding these counterparts is critical because a complete catalog will enable the study of energetics and emission mechanisms for all source populations within it. Moreover, blazars (AGN with relativistic jets pointed towards the observer at a viewing angle, $\theta_v < 10\degr$) detected above 10 GeV are powerful probes of the extragalactic background light \citep[EBL,][]{Dominguez15}, the integrated emission of all stars and galaxies in the Universe, which can shed insight into cosmological applications such as the measurement of the Hubble constant \citep{Dominguez2019}. However, that requires knowledge of their redshift. Associating the sources is thus the first step towards measuring their redshift and employing them for cosmological studies. Additionally, a complete 3FHL will be a critical resource for future observations with the upcoming Cherenkov Telescope Array \citep[CTA;][]{Hassan17, 4LAC2019}. \\
\indent One way to find  potential associations is by performing X-ray observations of the fields of $\gamma$-ray sources. The mechanisms responsible for creating the $\gamma$-ray emission in blazars, i.e., the synchrotron self-Compton process or the external Compton process, also emit in the X-rays \citep{Bottcher2007}. This is what motivates an X-ray search of $\gamma$-ray sources potentially associated to blazars. This X-ray radiation can localize the potential counterpart with greater reliability due to their $\sim$arcsecond positional uncertainties 
(see e.g. \citealp{Stroh13}\footnote{\url{https://www.swift.psu.edu/unassociated/}}, \citealp{SazParkinson16, Paiano17}). In addition, this improved positional localization enables the precise detection of the optical counterpart from the Ultra-Violet/Optical Telescope \citep[UVOT,][]{SwiftUVOT}, on board the Neil Gehrels \textit{Swift} Observatory \citep{Gehrels04}. Knowing the exact position will enable the follow up from ground based telescopes to measure the  redshifts of these sources.

\cite{Kaur19} have provided a likely association for 52 out of the 200 unassociated 3FHL sources  using the X-ray Telescope \citep[XRT,][] {SwiftXRT} also on board \textit{Swift}. This leaves $\sim$150 3FHL unassociated objects. Here, we follow the same approach and we analyze the {\it Swift} observations of 38 3FHL unassociated sources with the aim of identifying potential counterparts and understanding their nature. A machine learning algorithm is used here, for the sources which are believed to be associated to blazars, to understand whether they are flat-spectrum radio quasars (FSRQs, i.e., blazars with optical emission lines of equivalent width $>$ 5\AA) or BL Lacertae objects (BL Lacs, i.e., blazars with no emission lines in their optical spectrum) according to their spectral properties, such as the spectral photon index, color differences and variability.

 This paper is organized as follows: $\S$\ref{sec:data} discusses the multiwavelength data acquisition and analysis. The results from this analysis are reported in $\S$\ref{sec:res}. Then in $\S$\ref{sec:mach}, we describe our machine learning algorithm to classify these sources into BL Lacs and FSRQs and the corresponding results. Finally, $\S$\ref{sec:conc} contains the discussion and conclusions based upon our analysis.\\

 
\section{Data} \label{sec:data}
\subsection{Observations}
\indent \indent As part of the \textit{Swift} guest investigator cycle 14 (proposal 1417063, PI: Ajello\footnote{\url{https://swift.gsfc.nasa.gov/proposals/c14\_acceptarg.html\#abstracts}}), \textit{Swift}-XRT observed 20 bright unassociated 3FHL sources without any previous X-ray observation. Then, we cross-matched the remaining 119 unassociated and 9 unknown class sources in the 3FHL catalog with archival \textit{Swift} observations. We found 97 unassociated 3FHL sources that had been observed with \textit{Swift}-XRT. We selected observations where the 3FHL source fell within 20$\arcmin$ of the XRT pointing and had an XRT exposure $> 2$ ks to ensure reasonable statistics. This left us with 18 additional sources. For each target, we stacked all the XRT exposures found in the archive.

\subsection{Swift XRT Data Analysis}
\indent The analysis was performed using HEASARC version 6.26.1\footnote{\url{https://heasarc.gsfc.nasa.gov/docs/software.html}} and XSPEC version 12.10.1\footnote{\url{https://heasarc.gsfc.nasa.gov/xanadu/xspec/}} for the spectral fitting. The source spectra were extracted using a circular region with a radius ranging from 10$\arcsec-$15$\arcsec$ depending on the brightness of the source. The background spectra were obtained from an annular region centered on the source with inner radius 35.4$\arcsec$ and outer radius 70.7$\arcsec$. All spectra were fit in the 0.3$-$10\,keV regime with the Tuebingen-Boulder ISM absorption model \citep[\texttt{tbabs},][]{Wilms2000}. The Galactic column densities in the direction of the sources were determined following \cite{Kalberla05}. Most spectra were binned with 3 counts per bin while the remaining five sources were bright enough to use 10 counts per bin. Spectral fitting was performed with C-statistic for the low-count sources and $\chi^2$ statistics for the remaining\footnote{If c-stat was used, the results are in agreement.}. The  parameters of all X-ray sources are reported in  Table \ref{tab:xrt}. 

\subsection{Swift-UVOT Data Analysis}
\indent All of the sources observed by XRT also had an observation conducted in at least one \textit{Swift}-UVOT filter (except 3FHL J0737.5+6534 and 3FHL J1907.0+0713). The data were downloaded from the HEASARC archive and each cleaned sky image was loaded into DS9. A 5$\arcsec$ circular region was created at the position of the XRT source. In some cases, the UVOT counterpart was not centered in the XRT region, so this circle was moved slightly (no more than 3$\arcsec$) to enclose the entire source. 3FHL J1439.9$-$3955, 3FHL J1719.0$-$5348, and 3FHL J2030.4$+$2236 required a 4$\arcsec$ region to eliminate any overlap from a bright source nearby. Since the UVOT fields for these sources were crowded, it was not possible to select the usual annular background region around the source. Instead circular regions of radii 20$\arcsec$ were selected for the background from within the field where no other source was present. We provide AB magnitudes for all the detected UVOT counterparts of the XRT sources. 3FHL J1405.1$-$6118 is not included in the results due to an extremely high extinction value of A$_V$ = 67.8\footnote{A$_V$ represents the extinction in the V band.}. The results of the analysis are listed below in Table \ref{tab:uvot}. 

\subsection{Archival Data}
The NASA/IPAC Extragalactic Database (NED)\footnote{\url{https://ned.ipac.caltech.edu/simplesearch}} and SIMBAD\footnote{\url{http://simbad.u-strasbg.fr/simbad/}} were used to provide information at lower energies about the X-ray sources adopting a search radius of 5$\arcsec$. These results can be found in Table \ref{tab:count}. \\

\section{Results} \label{sec:res}

In the sample of 38 unassociated 3FHL sources, 22 contained at least one X-ray object in the field. Of these 22, 11 were high-latitude $(|b| > 10 \degr)$ and 11 were low-latitude $(|b| \leq 10 \degr)$. As discussed in $\S$\ref{sec:chance}, it is highly unlikely these sources are chance coincidences. Two of the 3FHL sources had two X-ray objects within 4$\arcmin$ of the 95\% confidence region boundary, leaving us with 24 X-ray sources to analyze. 

\begin{figure*} [h!]
\hbox{
\hspace{-2cm}
\includegraphics[scale=0.43]{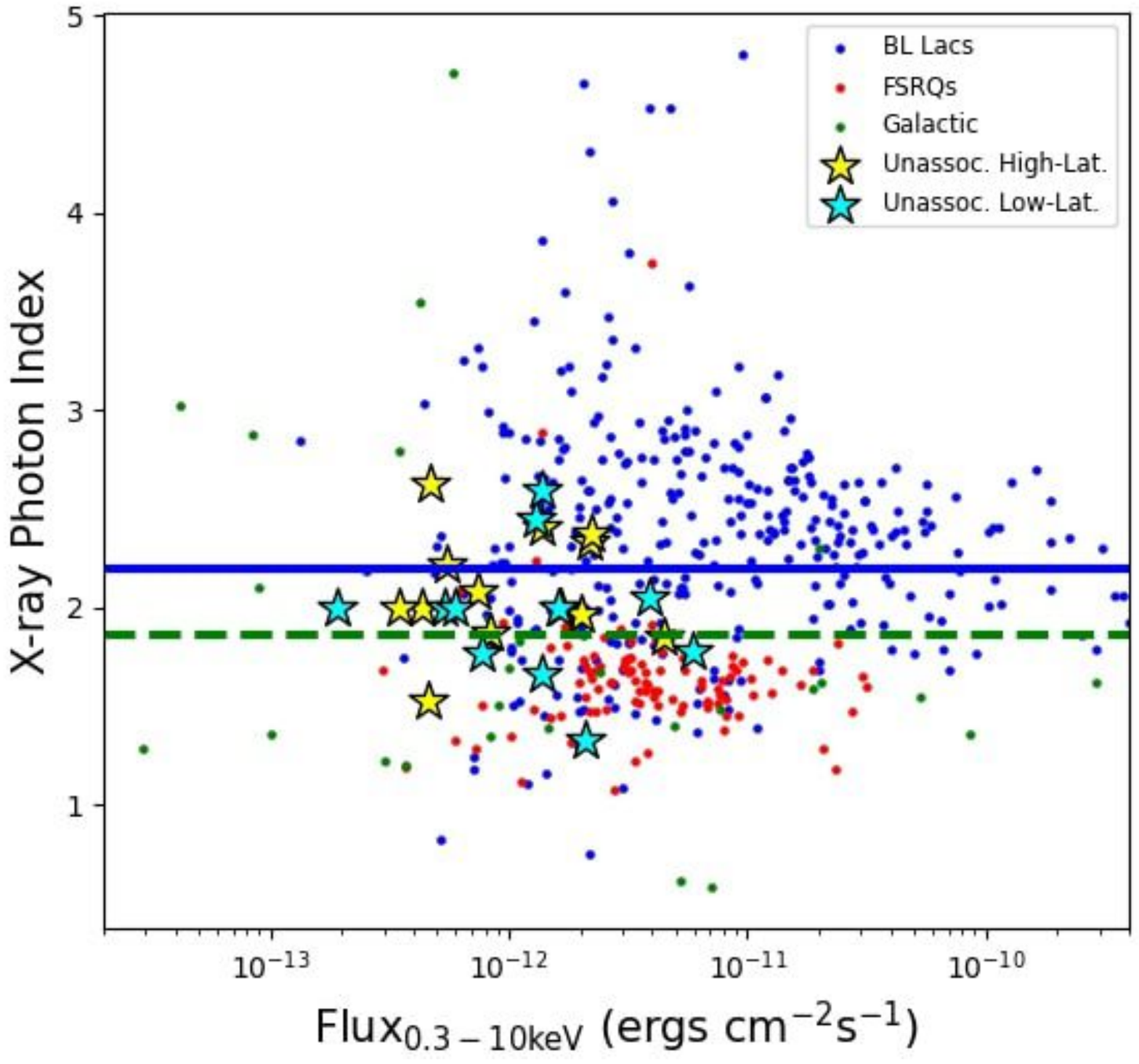}
\hspace{-2cm}
\includegraphics[scale=0.57]{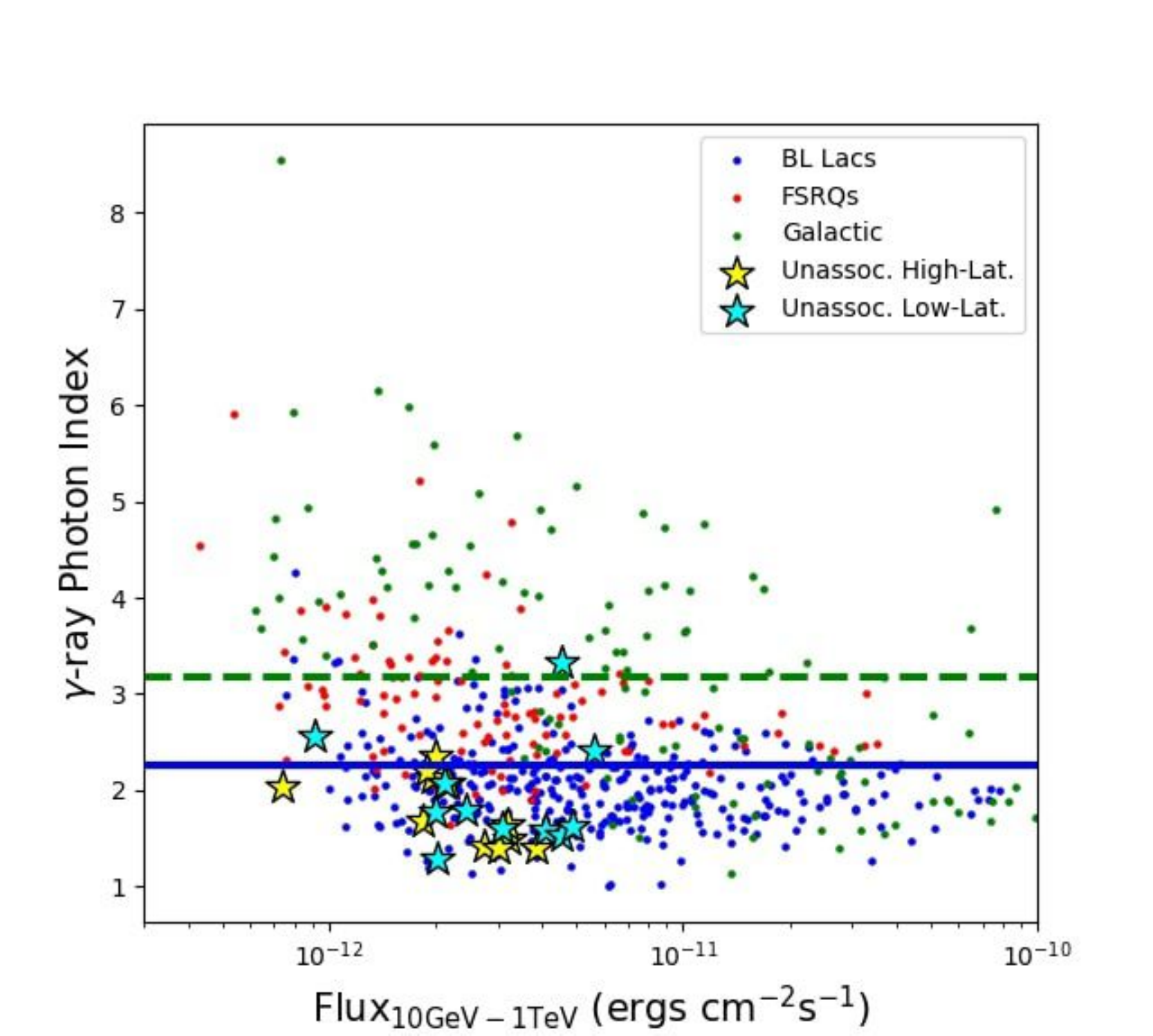}
}
\caption{The left image shows the distribution of X-ray photon indices versus XRT fluxes (0.3$-$10\,keV) and the right shows the $\gamma$-ray photon indices versus the 3FHL fluxes (10\,GeV$-$1\,TeV). The red circles represent known FSRQs, the blue circles represent known BL Lacs, and the green known Galactic sources. The blue and green lines represent the average photon index value for blazars and Galactic sources respectively. The yellow stars are the high-latitude sources in our sample and the cyan stars are the low-latitude sources. We note that BCUs from the 3FHL are mostly found in the locus occupied by FSRQs and BL Lacs. }
\label{fig:x+g}
\end{figure*}

\begin{figure}
    \centering
    \includegraphics[scale=0.4]{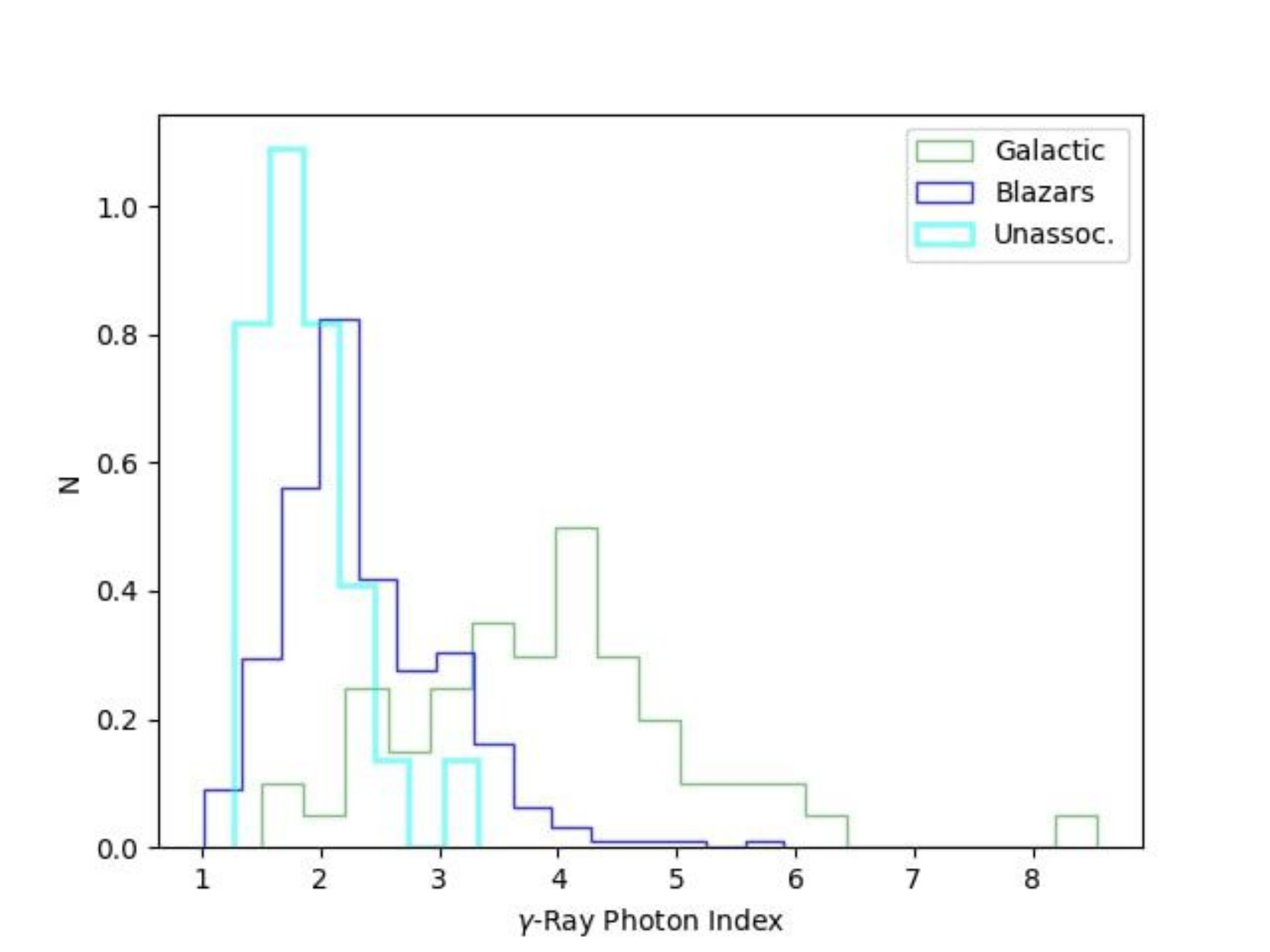}
    \caption{The normalized distributions of photon indices for blazars, Galactic sources, and the sample of unassociated sources with a $\gamma$-ray flux $<$ 7 $\times10^{-12}$\,erg cm$^{-2}$ s$^{-1}$.}
    \label{fig:g_spec}
\end{figure}

\begin{figure*} [htp]
    \centering
    \includegraphics[scale=0.5]{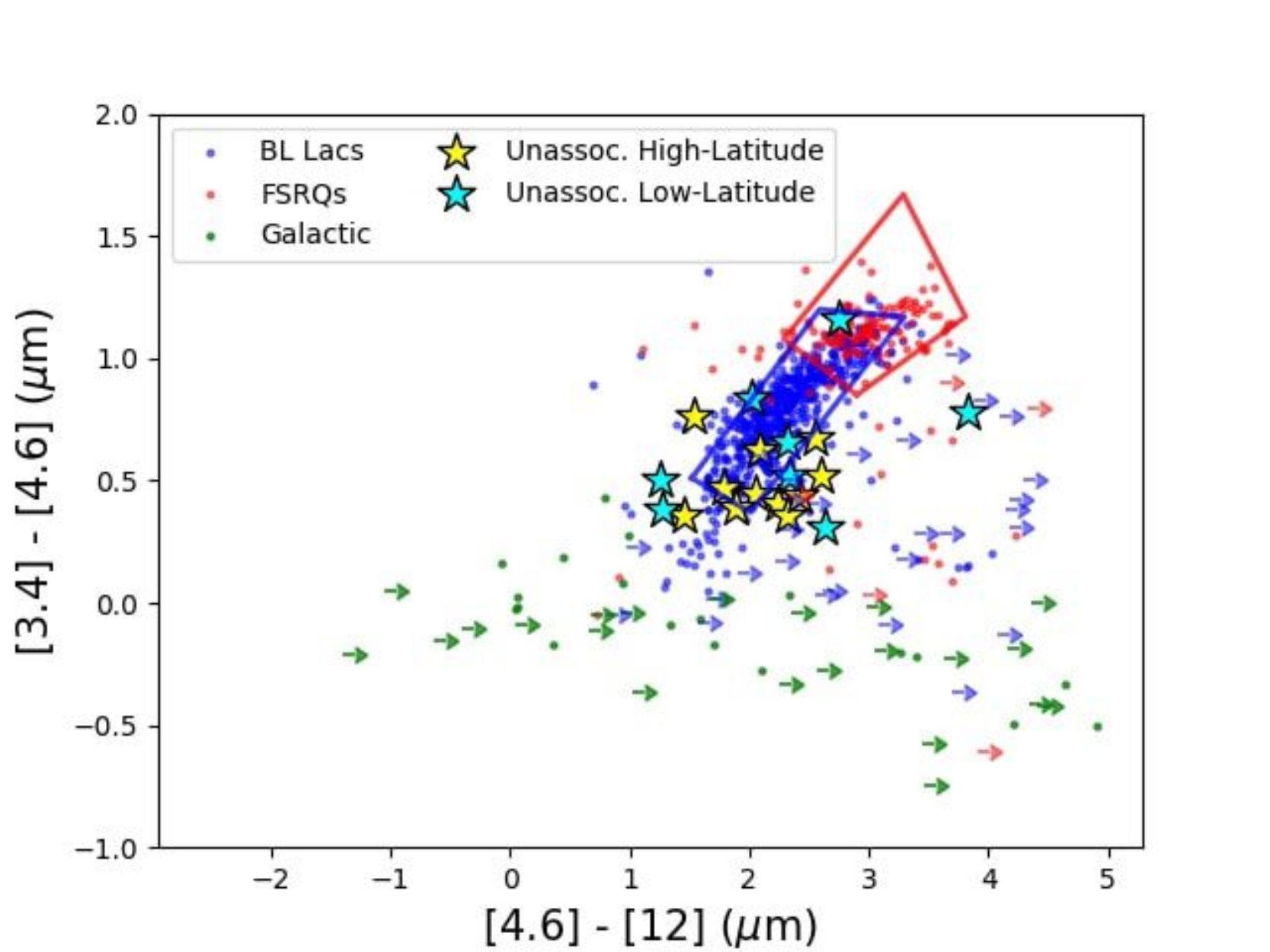}
    \caption{The \textit{WISE} blazar strip in the color space w1 (3.4$\mu$m) $-$ w2 (4.6$\mu$m) vs w2 $-$ w3 (12$\mu$m).
    The blue and red circles represent 439 known BL Lacs and FSRQs used in our machine learning algorithm while known Galactic sources are in green. The coordinates for the strip can be found in \cite{Massaro12}. The arrows represent sources with only an upper limit on the w3 magnitude. Two Galactic sources with a w1$-$w2 value $<$ -1 are not shown to improve readability.}
    \label{fig:bla_wise}
\end{figure*}

Approximately 80$\%$ of the objects in the 3FHL catalog are associated with blazars (FSRQs, BL Lacs, or blazar candidates of uncertain type, BCUs) with this fraction increasing to $\sim$90\% if low-latitude sources are excluded. Considering all the unassociated sources we analyzed have a 3FHL photon index between 1.2 and 3.5 and an energy flux between 0.5 $\times10^{-12}$ erg cm$^{-2}$ s$^{-1}$ and 7.5$\times10^{-12}$ erg cm$^{-2}$ s$^{-1}$, we calculated the fraction of sources classified as blazars in the 3FHL with a photon index and energy flux in that range. Approximately 96\% are blazars independent of Galactic latitude, however, if we only consider high-latitude sources, this increases to 98\%. \\
\indent We also compared the unassociated sources' multiwavelength data with that of classified blazars and Galactic objects. Figure \ref{fig:x+g} displays the XRT flux (0.3$-$10\,keV) and 3FHL flux (10\,GeV$-$1\,TeV) vs the photon index for the 24 analyzed sources, 439 3FHL sources classified as BL Lacs or FSRQs, and Galactic 3FHL sources. There is not a large discrepancy in the left plot with an average blazar X-ray photon index of 2.19 $\pm$ 0.60 compared to the average Galactic X-ray photon index of 1.86 $\pm$ 0.79, with the unassociated sources seemingly distributed evenly around both. However, the right plot clearly shows blazars with a harder photon index than Galactic objects, 2.25 $\pm$ 0.63 compared to 3.18 $\pm$ 1.29, and the 22 unassociated sources analyzed are even harder than the average blazars, aligning more with BL Lacs. This trend is further elucidated in Figure \ref{fig:g_spec} as the distribution of the photon spectral indices of unassociated sources is more aligned with the distribution of blazars than that of Galactic sources.
Moreover, Figure \ref{fig:bla_wise} displays a region of the infrared color-color space known as the \textit{WISE} blazar strip \citep{Massaro12}. Galactic objects are much more likely to have a w2 (4.6 $\mu$m) magnitude greater than or equal to w1 (3.4 $\mu$m) while blazars and the unassociated sources exhibit the reverse. Again, the unassociated sources most frequently fall into/near the BL Lac region which agrees with the fact that BL Lacs populate 80\% of the classified extragalactic sources in the 3FHL. Based on the above, the likely fraction of blazars among the 22 unassociated 3FHL sources analyzed here is 19/22. Of the 11 high-latitude and 11 low-latitude, 10 and 9 are likely to be blazars, respectively. While some of these sources do not have radio data available, the properties described above indicate a blazar nature.\\
\indent The three sources not believed to be blazars are 3FHL J0737.5$+$6534, 3FHL J1405.1$-$6118, and 3FHL J1907.0$+$0713, all of which are described as pulsar-like candidates in \cite{Hui20}. More specifically, \cite{Hui20} identifies 3FHL J1405.1$-$6118 as a new $\gamma$-ray binary. In addition to being a potential pulsar, 3FHL J1907.0$+$0713 is at low Galactic latitude ($b$ = $-$0.14$\degr$) and has a photon index ($\Gamma_{\gamma}=3.3$) more typical of Galactic objects in the 3FHL.\\
\indent According to \cite{Abdollahi20} and \cite{Ajello2020}, 3FHL J0737.5$+$6534 is associated with the star-forming galaxy NGC 2403. However, \cite{Xi2020} believes the $\gamma$-ray emission originated from the supernova SN 2004dj. Our analysis finds an X-ray source 2.28$\arcmin$ away from the 3FHL region that is spatially coincident with a high mass X-ray binary in NGC 2403, RX J073655.7+653542. Due to these findings, we have excluded it from consideration as a blazar. \\
\indent Another important application of the study of 3FHL unassociated sources is the indirect detection of
dark matter. \cite{Coronado2019_1910} and \cite{Coronado2019_1906} accomplish this
through the use of archival Swift observations. We note that the only source shared between those two
works and our analysis is 3FHL J0359.4$-$0235. We believe this source is a blazar, not dark matter,
because its properties align well with known blazars as visible in Figures 1 and 2. \\
\indent For all sources in Table \ref{tab:xrt}, we report UVOT magnitudes and archival counterparts at different wavelengths in Tables \ref{tab:uvot} and \ref{tab:count}, respectively.

\subsection{4FGL-DR2 Associations}
According to the second data release of the 4FGL catalog  \citep[4FGL-DR2,][]{Abdollahi20}, 7 of our 22 unassociated 3FHL sources have new associations with at least an 85\% probability. Five of these associations are consistent with the X-ray sources reported in Table \ref{tab:xrt} (3FHL J0233.5$+$0657\footnote{Right source in Figure \ref{fig:J0233}.}, 3FHL J0933.5$-$5240, 3FHL J1439.9$-$3955, 3FHL J1917.9$+$0331\footnote{Left source in Figure \ref{fig:J1917}.}, 3FHL J2321.6$-$1618). The other two associations are discussed in more detail below.
Of the 7 new associations, five are classified as BCUs while two are still of unknown class (3FHL J1917.9$+$0331 and 3FHL J1927.5$+$0153), meaning that the counterpart is also an unassociated source. All seven of these sources have been considered as blazars in this work and were included in our machine learning classification. Considering how the spectral properties of 3FHL J1917.9$+$0331 and 3FHL J1927.5$+$0153 align well with blazars, we included them in our machine learning algorithm despite having an unknown classification in the 4FGL-DR2. \\
\subsubsection{3FHL J0648.3+1744}
The 4FGL-DR2 reports GB6 J0648$+$1749 as the association with a 90\% probability, while the XRT source had no radio data available, and are thus not reported in Table \ref{tab:count}. GB6 J0648$+$1749 lies 1.2$\arcmin$ outside of the 3FHL 95\% confidence region whereas the XRT source is inside. Therefore, we believe the XRT source, SWIFT J064827$+$174423, is the more likely counterpart to the 3FHL source.

\subsubsection{3FHL J1927.5+0153}
NVSS J192729$+$015353 is reported as the association for 3FHL J1927.5$+$0153 in the 4FGL-DR2 with an 86\% probability whereas no radio data was available for the XRT source. Although both sources are inside the 3FHL 95\% confidence region, the XRT source is only 6$\arcsec$ away from the region's center while NVSS J192729$+$015353 is 34$\arcsec$ away. Furthermore, the XRT source has a potential WISE association that falls within the blazar strip while NVSS J192729$+$015353 has no WISE association. This is significant because the WISE data supports its blazar classification. For these reasons, we believe our XRT source is the more likely counterpart of 3FHL J1927.5$+$0153, yet it is possible these are all the same source if the NVSS positional uncertainty was underestimated. 

\subsection{Multiple X-Ray Sources} \label{mult:x-ray}
\begin{figure*} [htp]
\centering
\hbox{
\includegraphics[scale=0.3]{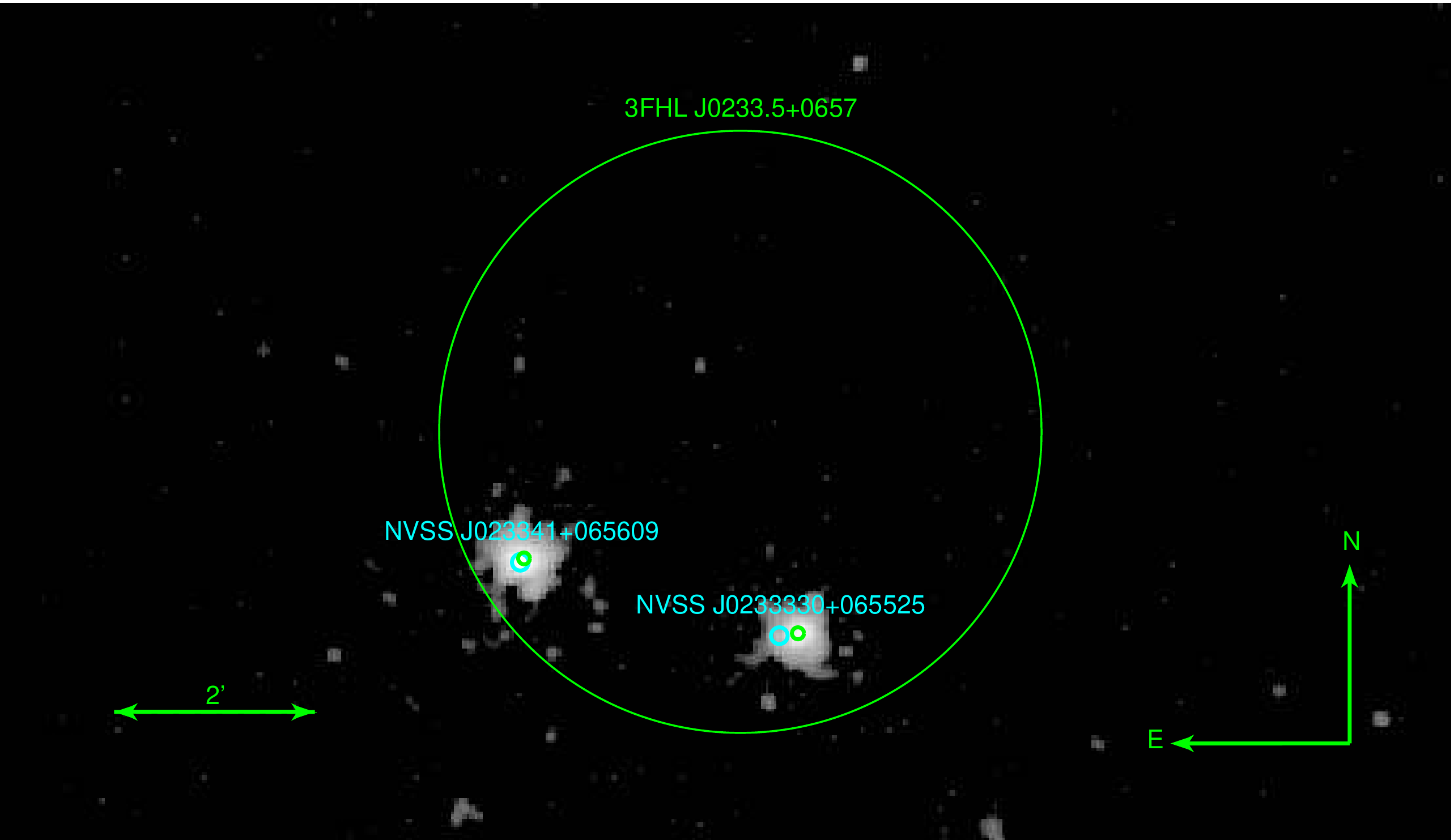}
\hspace{0.6cm}
\includegraphics[scale=0.3]{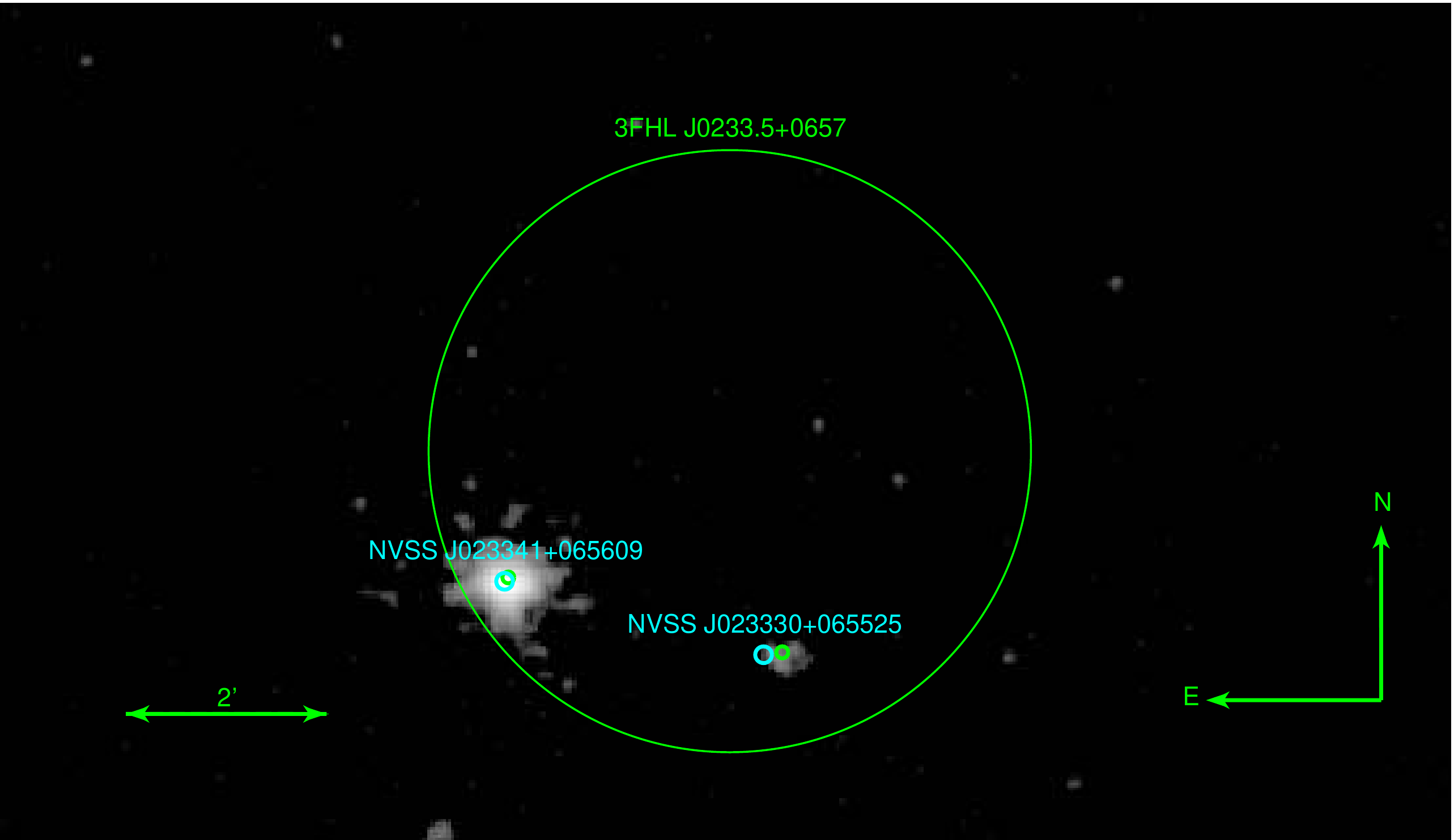}
}
\caption{0.3$-$10.0\,keV \textit{Swift}-XRT images of 3FHL J0233.5$+$0657. The left image was taken in June 2018 with an exposure of $\sim$3.1\,ks. The right image was taken in November 2018 as a part of the \textit{Swift} guest investigator cycle 14 and has an exposure time of $\sim$3.7\,ks. 
The green circle in both images represents the 95\% confidence interval from the 3FHL catalog. A source appears in the left image that is barely visible in the right, suggesting it is variable. The small green circles represent the XRT coordinates from Table \ref{tab:xrt} with their associated uncertainties. The NVSS counterparts are given an uncertainty radius of 5\arcsec to be visible, when their actual size is $\sim$0.8\arcsec.}
\label{fig:J0233}
\end{figure*}

3FHL J0233.5$+$0657 and 3FHL J1917.9$+$0331 are the two sources with multiple X-ray sources. The XRT field of 3FHL J0233.5$+$0657 can be viewed in Figure \ref{fig:J0233}. Since it is high-latitude, we aimed to discover which of these two X-ray sources most likely matched our above conclusion of being a blazar. We found that both had a radio counterpart, NVSS J023341$+$065609 (left) and NVSS J023330$+$065525 (right), which is often expected because blazars emit in radio through synchrotron radiation \citep{Bottcher2007}. Next, both displayed very similar SEDs with the two-hump spectrum indicative of blazars. Finally, both are considered radio loud, i.e. a radio flux density to optical flux density ratio\footnote{Radio flux at 5\,GHz and optical flux in the B band.} $>$ 10, with a ratio of 314 $\pm$ 14 (left) and 438 $\pm$ 4 (right). \\
\indent With all those results being so similar, we looked into the sources' variability in an effort to understand their nature. The left image in Figure \ref{fig:J0233} shows the field during the $\sim$3.1\,ks taken from XRT and the right five months later. As is evident, the right source decreases significantly in flux (2.67 $\times10^{-12}$\,erg cm$^{-2}$ s$^{-1}$ to 0.23 $\times10^{-12}$\,erg cm$^{-2}$ s$^{-1}$). This increases its likelihood of being a variable blazar. However, because the properties of both sources are blazar-like, it is impossible to confidently derive which is the likely counterpart of 3FHL J0233.5$+$0657.

\begin{figure} [htb]
\centering
\includegraphics[trim={0 0 5cm 0},clip, scale=0.35]{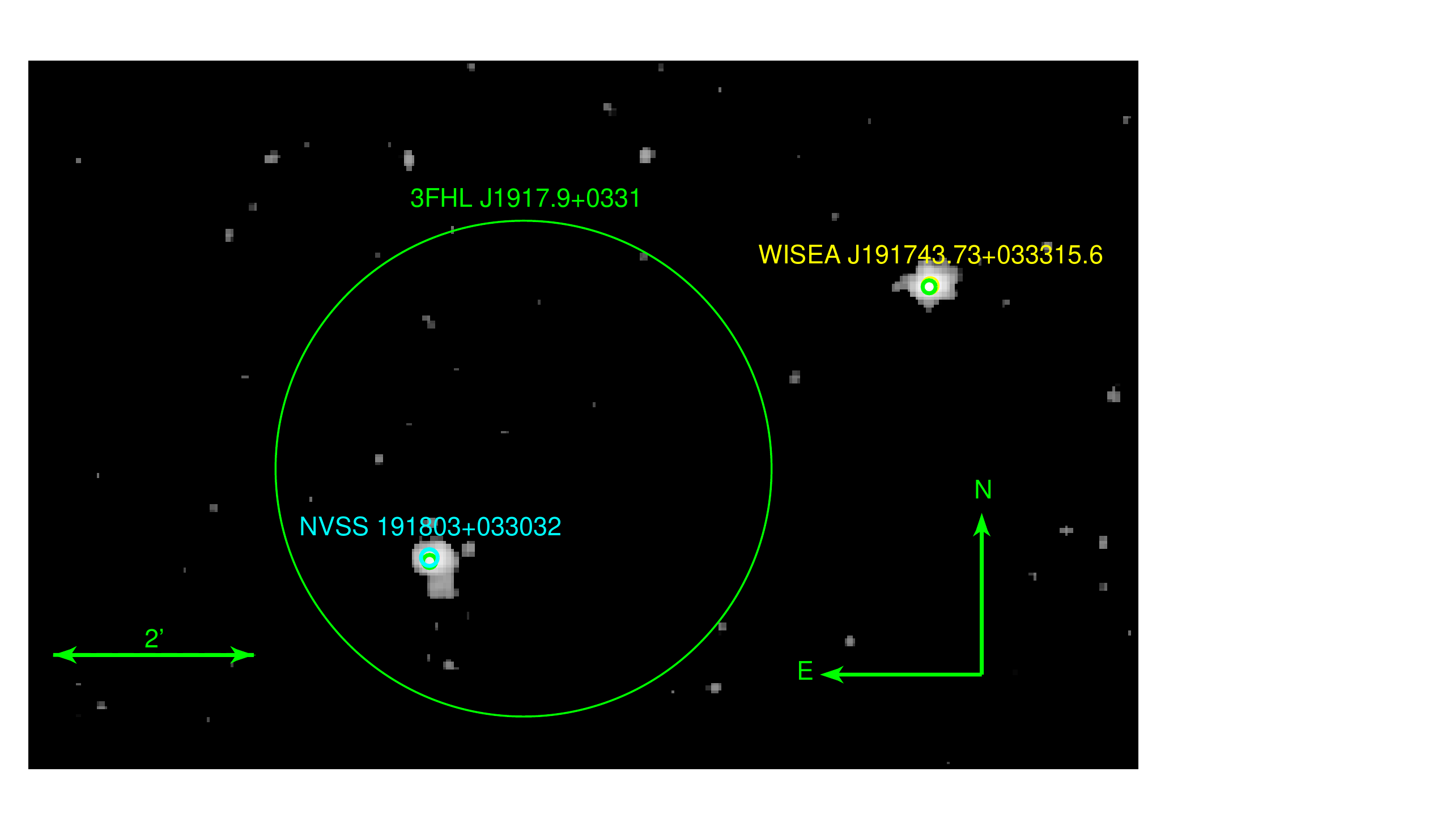}
\caption{0.3$-$10.0\,keV \textit{Swift}-XRT image of 3FHL J1917.9$+$0331. This image is comprised of three exposures all taken in September 2017 and totals up to $\sim$8\,ks. The green circle represents the 95\% confidence interval from the 3FHL catalog. The NVSS counterparts are given an uncertainty radius of 5\arcsec to be visible, when their actual size is $\sim$0.7\arcsec.}
\label{fig:J1917}
\end{figure}

The source 3FHL J1917.9$+$0331 also had two bright (Flux$_{0.3-10 keV} >$ 10$^{-12}$ erg cm$^{-2}$s$^{-1}$) X-ray sources in the field as seen in Figure \ref{fig:J1917}. While being inside the 3FHL region makes the left source the more likely counterpart, we wanted to verify it with additional information. Our analysis revealed the left source has an X-ray photon index of $2.45 \pm 0.43$ and a flux$_{0.3-10 {\rm keV}}$ = 1.31$\times$10$^{-12}$ erg cm$^{-2}$s$^{-1}$, both consistent with blazars
as evident in Figure \ref{fig:x+g}. Moreover, the left source has WISE and radio counterparts while the right lacks radio data. Therefore, we conclude that the left X-ray source (SWIFT J191804$+$033030), which falls within the 95\,\% 3FHL error region, is the more likely X-ray counterpart for 3FHL J1917.9$+$0331.

\subsection{Chance Coincidence} \label{sec:chance}
Following the procedure laid out in \cite{Xi2020}, we used a Poisson distribution to determine the likelihood of finding another X-ray source of similar flux inside the 3FHL 95\,\% uncertainty region. We calculated the chance probability as:
\begin{equation}
    P_{ch} = 1 - exp[-\pi (R_0^2 + 4\sigma_\gamma^2) \Sigma(>F_{th})],
\end{equation}
where R$_0$ is the angular distance between the 3FHL source and the X-ray source, $\sigma_\gamma$ is the 95\,\% uncertainty radius of the 3FHL source, and $\Sigma$($>$F$_{th}$) is the surface density of X-ray sources with flux greater than F$_{th}$. Our sample has an average R$_0$ = 2$\arcmin$ and $\sigma_\gamma$ = 2.3$\arcmin$. Using the newly released 4XMM-DR9 \citep{Webb2020}, we calculated a lower limit source density of 2.2 degrees$^{-2}$ at high-latitudes and with a flux $>$ 2.25 $\times10^{-12}$ erg cm$^{-2}$ in the 0.2$-$12\,keV band. This value was extrapolated from the average flux calculated in the 0.3$-$10\,keV band observed by XRT. These values give an estimate for the probability of chance coincidence equal to 1.5 $\times10^{-5}$.

\begin{center}

\begin{longrotatetable}
\begin{deluxetable*}{c c c c c c c c c c c c} 
    \tablecaption{\textit{Swift}-XRT Analysis \label{tab:xrt}}
    \tabletypesize{\footnotesize}
    \tablehead{\colhead{3FHL} & \colhead{SWIFT Name$^a$} & \colhead{X-Ray R.A.} & \colhead{X-Ray Decl.} & \colhead{Exp. Time} & \colhead{$\Gamma_X$$^b$} & \colhead{$N_H$$^c$} & \colhead{Flux$^d$} & \colhead{RL$^e$} & \colhead{$\chi^2$/d.o.f.} & \colhead{In 95\%}\\ 
    \colhead{} & & \colhead{(hh:mm:ss)} & \colhead{($\degree$:':'')} & \colhead{(ks)} & \colhead{} & \colhead{(X $10^{22}$ $cm^{-2}$)} & \colhead{(cgs)} & \colhead{} & \colhead{}} 
    \movetabledown=10mm
    \startdata
J0233.5$+$0657 $^f$ & J023341$+$065611 & 02:33:40.91 & 06:56:11.31 & 14.1 & 1.85 $\pm$ 0.07 &	0.062 & 4.49$^{+0.14}_{-0.16}$ & 314 $\pm$ 14 & 99/91 & Yes \\
J1439.9$-$3955 & J143951$-$395517 & 14:39:50.89 & $-$39:55:17.19 &	7.5 & 2.33 $\pm$ 0.15 & 0.066 & 2.17$^{+0.10}_{-0.11}$ & 78 $\pm$ 4 & 18.53/25 & Yes \\
J1451.8$-$4145 & J145149$-$414525 & 14:51:49.40 & $-$41:45:24.57 & 7.1 & 2.40 $\pm$ 0.22 & 0.077 & 1.37$^{+0.12}_{-0.84}$ & ... & 11.51/11 & Yes \\
J1719.0$-$5348 & J171856$-$535043 & 17:18:56.48 & $-$53:50:42.64 & 7.1 & 1.78 $\pm$ 0.13 & 0.145 & 5.9$^{+0.21}_{-0.30}$ & 29 $\pm$ 4 & 34.4/40 & Yes \\
J2321.6$-$1618 & J232137$-$161927 & 23:21:36.82 & $-$16:19:26.65 & 3.7 & 2.37 $\pm$ 0.23 & 0.018 & 2.22$^{+0.18}_{-0.16}$ & ... & 13.35/15 & Yes \\
\hline \\
3FHL & SWIFT Name & X-Ray R.A. & X-Ray Decl. & Exp. Time & $\Gamma_X$ & $N_H$ & Flux & RL & Cstat/d.o.f. & In 95\% \\ 
& & (hh:mm:ss) & ($\degree$:':'') & (ks) & & (X $10^{22}$ $cm^{-2}$) & (cgs) & \\ 
\hline \hline
J0057.9$+$6325 & J005758$+$632637 & 00:57:58.4 & 63:26:37.34 & 8.7 & 2.05 $\pm$ 0.24 & 0.81 & 3.86$^{+0.17}_{-0.22}$ & 24 $\pm$ 4$\ast$ & 70.1/54 & Yes \\
J0233.5$+$0657 $^g$ & J023330$+$065526 & 02:33:29.89 & 06:55:26.44 & 14.1 & 1.96 $\pm$ 0.13 & 0.062 & 2.01$^{+0.12}_{-0.10}$ & 438 $\pm$ 4$\ast$ & 100.1/105 & Yes \\
J0359.4$-$0235 & J035923$-$023459 & 03:59:23.39 & $-$02:34:59.46 & 6.1 & 2.00$\dag$ & 0.11 & 0.43$^{+0.08}_{-0.07}$ & 96 $\pm$ 6$\ast$ & 2.9/5 & Yes \\
J0528.4$+$3851 & J052831$+$385200 & 05:28:31.27 & 38:51:59.52 & 3.8 & 1.32 $\pm$ 0.81 & 0.51 & 2.08$^{+0.45}_{-0.52}$ & 61 $\pm$ 7$\ast$ & 5.0/10 & Yes \\
J0648.3$+$1744 & J064827$+$174423 & 06:48:26.67 & 17:44:23.19 & 9.7 & 2.00$\dag$ & 0.18 & 0.16$^{+0.03}_{-0.03}$ & ... & 0.6/2 & Yes \\
J0737.5$+$6534 & J073655$+$653530 & 07:36:54.99 & 65:35:30.08 & 25.1 &	1.52 $\pm$ 0.29 & 0.11 & 0.46$^{+0.09}_{-0.06}$ & ... & 31.0/32 & No, 2.28\arcmin $^h$ \\
J0933.5$-$5240 & J093334$-$524021 & 09:33:33.50 & $-$52:40:20.54 & 8.0 & 2.00$\dag$ & 0.84 & 0.51$^{+0.06}_{-0.07}$ & 6440 $\pm$ 890$\ast$ & 1.7/4 & Yes \\
J1405.1$-$6118 & J140514$-$611826 & 14:05:14.20 & $-$61:18:25.91 & 23.3 & 2.00$\dag$ & 1.8 & 0.84$^{+0.05}_{-0.04}$ & ... & 8.2/11 & Yes \\
J1855.3$+$0751 & J185520$+$075140 & 18:55:19.86 & 07:51:39.97 & 5.0 & 2.00$\dag$ & 0.72 & 1.28$^{+0.22}_{-0.20}$ & 42 $\pm$ 7$\ast$ & 4.7/3 & Yes \\
J1907.0$+$0713 & J190706$+$071953 & 19:07:06.23 & 07:19:52.97 & 14.4 & 1.76 $\pm$ 0.88 & 1.5 & 0.77$^{+0.04}_{-0.21}$ & ... & 4.8/7 & No, 3.74\arcmin \\
J1917.9$+$0331 $^i$ & J191804$+$033030 & 19:18:03.76 & 03:30:30.37 & 7.9 & 2.45 $\pm$ 0.43 & 0.39 & 1.31$^{+0.13}_{-0.08}$ & ... & 17.7/21 & Yes \\
J1917.9$+$0331 $^j$ & J191744$+$033315 & 19:17:43.77 & 03:33:14.98 & 7.9 & 1.66 $\pm$ 0.34 & 0.39 & 1.39$^{+0.20}_{-0.16}$ & ... & 21.8/25 & No, 1.75\arcmin \\
J1927.5$+$0153 & J192732$+$015355 & 19:27:31.59 & 01:53:54.62 &	3.8 & 2.59 $\pm$ 0.44 & 0.18 & 1.38$^{+0.18}_{-0.12}$ & ... & 14.5/15 & Yes \\
J2030.2$-$5037 & J203024$-$503411 & 20:30:24.18 &	$-$50:34:11.28 & 2.8 & 1.87 $\pm$ 0.40 & 0.024 & 0.84$^{+0.26}_{-0.09}$ & ... & 15.2/15 & No, 0.90\arcmin \\
J2030.4$+$2236 & J203031$+$223439 & 20:30:31.33 & 22:34:39.31 & 5.0 & 2.00$\dag$ & 0.14 & 1.00$^{+0.18}_{-0.26}$ & ... & 6.1/3 & Yes \\
J2104.5$+$2117 & J210416$+$211813 & 21:04:15.95 & 21:18:12.50 & 2.1 & 2.00$\dag$ & 0.10 & 0.30$^{+0.11}_{-0.05}$ & ... & 1.5/4 & No, 2.41\arcmin \\
J2105.9$+$7508 & J210606$+$750921 & 21:06:05.61 & 75:09:20.94 & 4.7 & 2.21 $\pm$ 0.93 & 0.14 & 0.55$^{+0.10}_{-0.08}$ & ... & 8.1/4 & Yes \\
J2159.6$-$4619 & J215936$-$461954 & 21:59:35.82 & $-$46:19:53.58 & 3.1 & 2.62 $\pm$ 0.36 & 0.014 & 0.47$^{+0.06}_{-0.05}$ & ... & 19.4/22 & Yes \\
J2239.5$-$2439 & J223928$-$243945 & 22:39:28.48 & $-$24:39:44.91 & 8.1 & 2.08 $\pm$ 0.27 & 0.015 & 0.74$^{+0.09}_{-0.08}$ & ... & 42.8/30 & Yes \\
    \enddata 
\vspace{0.5cm}
 \textbf{Notes.} $\dag$ designates the model fit did not yield well-constrained results. Therefore, the photon index was frozen at 2.00. \\ 
$^a$ Name we designate for the sources detected by XRT. \\
$^b$ X-ray photon index. \\ 
$^c$ Galactic column density. \\
$^d$ Unabsorbed flux in the 0.3$-$10 keV band ($\times10^{-12}$ erg cm$^{-2}$ s$^{-1}$). \\
$^e$ Radio Loudness, i.e., ratio of flux density at 5 GHz and flux density in B band. \\
$\ast$ Denotes only a lower limit for the B flux when calculating the radio loudness. \\
$^f$ Left source in Figure \ref{fig:J0233}. \\
$^g$ Right source in Figure \ref{fig:J0233}. \\
$^h$ Distance from boundary of 3FHL 95\% confidence region to center of X-ray source. \\
$^i$ Inside 3FHL in Figure \ref{fig:J1917}. \\
$^j$ Outside 3FHL in Figure \ref{fig:J1917}. 
\end{deluxetable*} 
  \end{longrotatetable}

\end{center}

\begin{deluxetable*}{c c c c c c c}
    \centering
   \tablecaption{\textit{Swift}-UVOT Magnitudes \hspace{7in}
    The sources in bold are high-latitude.}
    \label{tab:uvot}
    \tablehead{\colhead{Source Name} &  \colhead{W2} & \colhead{M2} & \colhead{W1} & \colhead{U} & \colhead{B} & \colhead{V}}   

\startdata
J0057.9$+$6325 & $> 11.65$ & $> 10.02$ &	$> 12.56$ & $> 13.80$ &	$> 14.01$ & $14.10 \pm 0.27$ \\
\textbf{J0233.5$+$0657}  &  $19.85 \pm 0.16$ & $20.04 \pm 0.29$ & $19.40 \pm 0.18$ & $19.08 \pm 0.19$ & $18.71 \pm 0.23$ & $> 18.48$\\
\textbf{J0233.5$+$0657} & $20.29 \pm 0.23$ &  $20.25 \pm 0.32$ & $19.68 \pm 0.23$ &  $19.00 \pm 0.18$ & $ > 19.15$ & $> 18.42$\\
\textbf{J0359.4$-$0235} & $> 19.36$ & $> 18.72$ & $>18.95$ & $> 18.69$ &  $> 18.06$ & $> 17.39$ \\
J0528.4$+$3851 & $> 11.65$ &	$> 9.92$ & $> 12.74$ & $> 14.31$ &  $> 14.55$ & $> 14.89$ \\
J0648.3$+$1744 & $> 20.43$ & $> 19.93$ & $19.78 \pm 0.36$ & 	$19.31 \pm 0.31$  & $> 18.79$ & $> 18.04$ \\
J0933.5$-$5240 & $> 7.69$ & $> 5.68$ & $> 9.84$ & $> 12.22$ &  $> 12.86$ & $> 13.57$ \\
\textbf{J1439.9$-$3955} & $19.66 \pm 0.13$ & $19.30 \pm 0.15$ & $18.94 \pm 0.13$ &  $18.43 \pm 0.10$ & $17.93 \pm 0.11$ & $17.54 \pm 0.14$ \\
\textbf{J1451.8$-$4145} & $19.96 \pm 0.21$  &	$19.68 \pm 0.25$ &	$19.89 \pm 0.28$ & $19.28 \pm 0.22$ & $18.34 \pm 0.18$ & $18.38 \pm 0.36$ \\
J1719.0$-$5348 & $> 19.76$ & $> 19.03$ & $> 19.33$ & $18.24 \pm 0.18$ & $16.82 \pm 0.10$ & $15.89 \pm 0.09$ \\
J1855.3$+$0751 & $> -7.61$ &	$> -12.12$ & $> -2.75$ & $> 2.72$  & $> 4.85$ & $> 7.45$ \\
J1917.9$+$0331 & & $> 16.67$ & $> 15.58$ & & & \\
J1917.9$+$0331 & & $> 15.40$ & $> 16.31$ & & & \\
J1927.5$+$0153 & & & & $17.25 \pm 0.09$ & & \\
\textbf{J2030.2$-$5037} & & $21.10 \pm 0.17$ & & & & \\
J2030.4$+$2236 & & & $19.82 \pm 0.22$ & $ 18.75 \pm 0.15$ & \\
\textbf{J2104.5$+$2117} & & &  $19.38 \pm 0.34$ & $18.69 \pm 0.21$ & \\
\textbf{J2105.9$+$7508} & $> 16.65$ & & & & & \\
\textbf{J2159.6$-$4619} & & & & $19.12 \pm 0.05$ & & \\
\textbf{J2239.5$-$2439} & & & $20.34 \pm 0.20$ & & & \\
\textbf{J2321.6$-$1618} & & & $18.35 \pm 0.06$ & & 
\enddata

\end{deluxetable*}

\newpage
\begin{deluxetable*}{c c c c c} 
    \centering
    \tablecaption{\textit{Multi-Wavelength Data} \label{tab:count}}
    \tablehead{\colhead{3FHL} & \colhead{Radio} &  \colhead{2MASS} & \colhead{WISE} & 
    \colhead{Ultraviolet}}   
    \startdata
J0057.9$+$6325 & NVSS J005758$+$632636 & J00575838$+$6326390 & J005758.38$+$632639.3	 & ... \\
J0233.5$+$0657 $^a$ & NVSS J023341$+$065609 & J02334098$+$0656114 & J023340.99$+$065611.1 & GALEX J023340.97$+$065611.4 \\
J0233.5$+$0657 $^b$ & NVSS J023330$+$065525 & J02332999$+$0655260 & J023329.97$+$065526.3	 & GALEX J023329.9+065528 \\
J0359.4$-$0235 & NVSS J035923$-$023501 & J03592349$-$0235022 & J035923.48$-$023501.8 & ... \\		
J0528.4$+$3851 & NVSS J052831$+$385156 & ... & J052831.61$+$385200.5	 & ... \\
J0648.3$+$1744 & ... & J06482642$+$1744235 & J064826.73$+$174422.5	 & ... \\		
J0737.5$+$6534 & ... & J07365473$+$6535277	 & ... & ... \\		
J0933.5$-$5240 & TGSSADR J093333.2$-$524020	& J09333316$-$5240192 & J093333.17$-$524019.3	 & ... \\		
J1405.1$-$6118 & ...	& J14051441$-$6118282 & J140514.40$-$611827.7	& ... \\		
J1439.9$-$3955 & NVSS J143951$-$395517 & J14395085$-$3955185 & J143950.86$-$395518.8 & ... \\
J1451.8$-$4145 & ... & J14514931$-$4145034 & J145149.34$-$414503.6 & ... \\		
J1719.0$-$5348 & MGPS J171855$-$535042 & J17185595$-$5350484 & ...  & ... \\		
J1855.3$+$0751 &	NVSS J185520$+$075140	& ... & ... & ... \\	
J1907.0$+$0713 & ...	& J19070619$+$0719545 & ... & ... \\		
J1917.9$+$0331 $^c$ &	NVSS J191803$+$033032 & J19180361$+$0330300 & J191803.60$+$033031.1	 & ... \\		
J1917.9$+$0331 $^d$ & ... & ... &	 J191743.73$+$033315.6 & ... \\	
J1927.5$+$0153 & ... & ... & J192731.14+015357.9 & ...	 	 \\		
J2030.2$-$5037 & TGSSADR J203023.9$-$503411 & ... & J203024.04$-$503413.0 & GALEX J203024.0-503413
 \\		
J2030.4$+$2236 & NVSS J203031$+$223439	& ... & ... & ... \\		
J2104.5$+$2117 & NVSS J210415$+$211805 & ... & J210415.92$+$211808.2	& ... \\	
J2105.9$+$7508 & NVSS J210606$+$750926 & ... & J210605.46$+$750920.7	 & ... \\	
J2159.6$-$4619 & ... & ... & J215936.14$-$461953.9 & ... \\
J2239.5$-$2439 & NVSS J223928$-$243943 & J22392882$-$2439441 & J223928.82$-$243944.2 & ...\\
J2321.6$-$1618 & NVSS J232137$-$161935 & J23213700$-$1619282 & J232136.98$-$161928.3	 & GALEX J232136.9-161928
\enddata
\vspace{0.3cm}
$^a$ Left source in  Figure \ref{fig:J0233}. \\
$^b$ Right source in Figure \ref{fig:J0233}. \\
$^c$ Inside 3FHL in Figure \ref{fig:J1917}. \\
$^d$ Outside 3FHL in Figure \ref{fig:J1917}. 

\end{deluxetable*}

\section{Classification with Machine Learning} \label{sec:mach}
\indent Generally, multiwavelength analysis is necessary to accurately classify every \textit{Fermi}-LAT detected source. However, collecting all the necessary data is highly time-intensive, leading to a growing number of unidentified sources in each \textit{Fermi}-LAT catalog release. Despite this challenge, the majority of sources in the catalog are associated with blazars. Given that 19 of our sources are consistent with blazars as reported in $\S$\ref{sec:res}, we proceeded to determine their blazar type. In order to classify each source, we incorporated five parameters (displayed in Table \ref{tab:param}) into our algorithm with the ability to differentiate BL Lacs from FSRQs. These parameters were taken from the 3FHL, 1SXPS \citep{Evans14}, and ALLWISE catalogs \citep{Cutri13}, and are described further in  $\S$\ref{sec:sample}. Several machine-learning techniques have been successful in their classification of \textit{Fermi}-LAT unidentified sources, e.g., \cite{Ackermann12}, \cite{Mirabal12}, \cite{Mirabal16}, \cite{SazParkinson16}, \cite{Salvetti17}, and \cite{Kaur19}. In this work, we employed two of the most commonly used methods: Decision Tree \citep[DT;][]{Quinlan90} and Random Forest \citep{Breiman01}. 

\subsection{Decision Tree}
\indent \indent The DT classifier is a supervised machine-learning algorithm that uses particular parameters based on the input data to split it into two or more categories. The data is continually split into branches and nodes and concludes only when each data point is designated to one of the categories. The nodes are split using the Gini index, an impurity measurement, which determines the parameter that can best separate the data, thus maximizing the accuracy of classification. The Gini impurity parameter states how likely it is to improperly label a data point. The DT algorithm reduces this value by splitting the sample into branches until this index reaches its minimum of zero. The index is defined as 
\begin{equation}
    G = 1 - \sum_{i = 1}^{J} p_i ^2,
\end{equation}
where J is the total number of categories and $p_i$ is the fraction of items labeled with category $i$ in the data sample. A larger G value indicates a greater level of disparity between the two classes for a given parameter. The DT is split until G reaches zero, thus assigning each data set a class along the way. A data set with known classification is split into two groups: one to train the classifier and the other to test its accuracy. Then, the classifier can assess an unclassified data set.

\subsection{Random Forest}
\indent \indent The random forest method is the most commonly used supervised machine-learning technique for both classification and regression. This works as an ensemble algorithm following the same principles as a DT classifier. In a random forest classifier, numerous DT algorithms are run with each assigning a class to every data point. The combined result of these predicted classes is the final result for each source in the data set. The random forest is preferred over a single DT because it solves the problem of overfitting \citep{Hastie09}. This method was used to classify each source in our sample as a BL Lac or an FSRQ with an associated probability.

\subsection{Sample and Parameter Selection} \label{sec:sample}
\indent \indent We utilized the DT and Random Forest classifiers implemented in \textit{sklearn0.20.0} library \citep{Pedregosa12} in \textit{python 2.7} on a sample of 439 3FHL blazars (336 BL Lac objects and 103 FSRQs). This sample was chosen from the 3FHL catalog because they possess known values for all five parameters listed in Table \ref{tab:param}. Our goal was to follow the process established in \cite{Kaur19} and include the 3FGL Index as the sixth parameter, but 16 of our 24 unassociated X-ray sources did not have a 3FGL photon index. We found 22 of our 24 X-ray sources had a 4FGL photon index, but including it in our machine learning produced identical classifications with similar probabilities. Therefore, we did not include the 4FGL photon index in our algorithm. The five parameters chosen have been shown to distinguish BL Lacs from FSRQs. Generally, BL Lac objects exhibit harder spectra in Gamma rays \citep[e.g.,][]{Abdo10, Ackermann15} and softer in X-rays \citep[e.g.,][]{Donato01} when compared to FSRQs as is visible in Figure \ref{fig:x+g}. Therefore, we selected the $\gamma$-ray photon index from the 3FHL catalog and the X-ray photon index from the 1SXPS catalog. Moreover, \cite{Massaro12} introduced a method to classify blazars of unknown type using a four-filters WISE color-color diagram \citep[also used in][]{DAbrusco14}. From their diagram and Figure \ref{fig:bla_wise}, it can be seen that BL Lac objects occupy the bluer region of the parameter space. Finally, FSRQs exhibit more variability than BL Lacs, which is quantified by the Variability Bayesian Blocks in the 3FHL. The values range from 1 to 15, where 1 implies no variability and 15 implies high variability. 20\% of the sample was separated into a test set with the remaining 80\% being used to train the classifiers. 

\begin{deluxetable}{c c c}
    \centering
    \tablecaption{Parameters for Blazar Classification \label{tab:param}}
    \tabletypesize{\scriptsize}
    \movetableright=-12mm
    \tablehead{\colhead{Parameter} &  \colhead{Catalog} & \colhead{References}} 
    \startdata
         X-ray Photon Index & Table \ref{tab:xrt} for unknown sample  & See Table \ref{tab:xrt}\\
         & 1SXPS for training set & \cite{Evans14} \\
         Variability Bayesian Blocks & 3FHL & \cite{Ajello17} \\
         w1$-$w2 & AllWISE & \cite{Cutri13} \\
         w2$-$w3 & AllWISE & \cite{Cutri13} \\
         Gamma-ray Photon Index & 3FHL & \cite{Ajello17} \\
    \enddata
\end{deluxetable}

\begin{figure}
    \centering
    \includegraphics[scale=0.4]{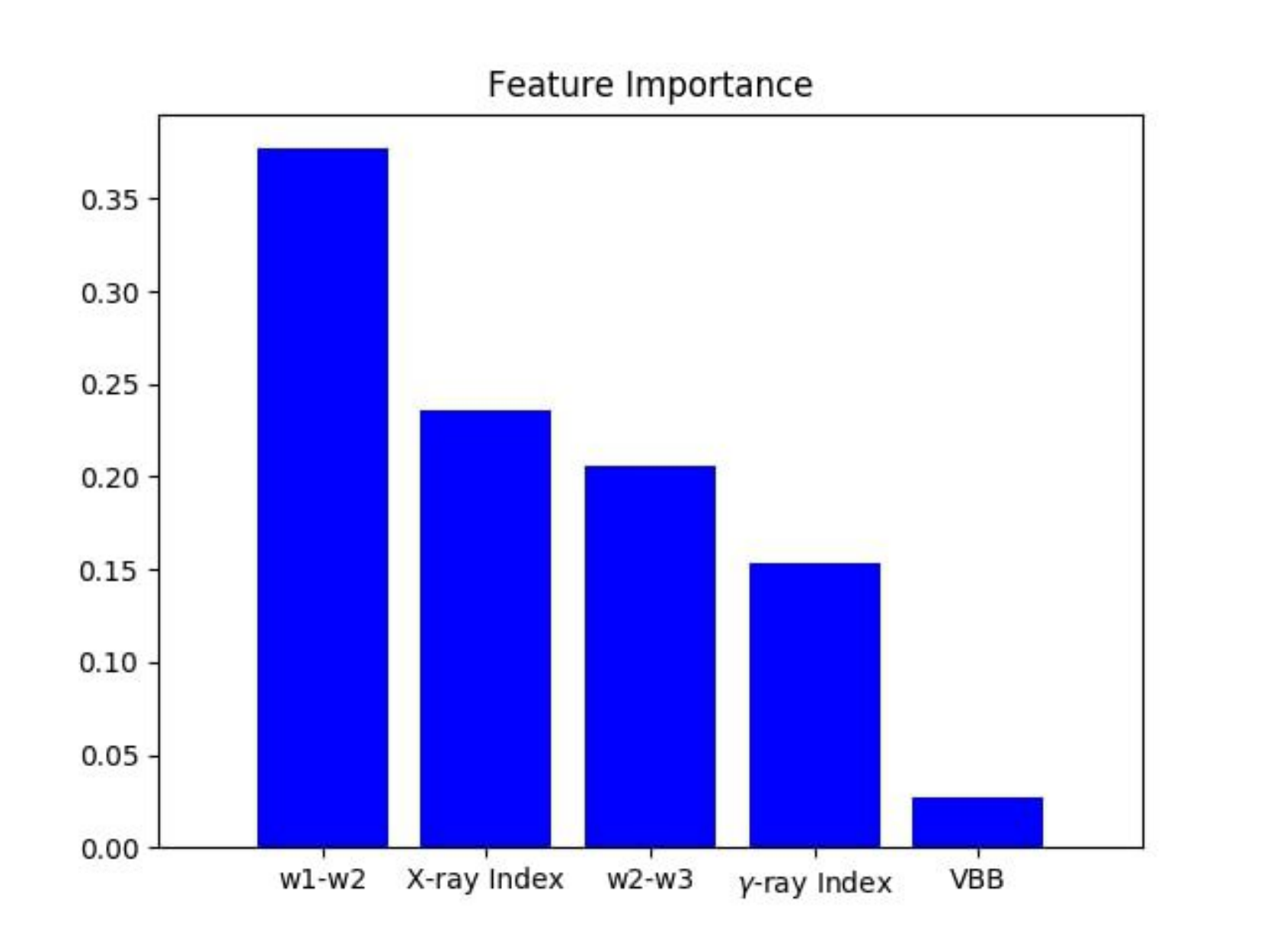}
    \caption{The feature importance for each parameter in our Random Forest classifier. Here "VBB" represents the parameter Variability Bayesian Blocks.}
    \label{fig:feat_im}
\end{figure}

\subsection{Results} \label{sec:mach_res}
\indent \indent Of the 19 unidentified 3FHL sources believed to be blazars, three of them could not be included in our algorithm because they did not have a WISE counterpart listed in Table \ref{tab:count}. Including 3FHL J0233.5$+$0657 with two X-ray objects, we had 17 sources in our algorithm. The feature importance, i.e., a score that expresses the relative importance of each parameter used in the classification, for our five parameters is shown in Figure \ref{fig:feat_im} while results from the machine learning classification algorithms are displayed in Table \ref{tab:mach}. Figure \ref{fig:feat_im} clearly indicates that the WISE colors w1$-$w2 had the greatest impact on classifying our sample while the Variability Bayesian Blocks had the least due to the small differences in variability between the 17 sources. We employed our DT classifier on the test data set (67 BL Lac objects and 21 FSRQs) and yielded an accuracy of 84\%. When this classifier was applied to the unclassified sample of 17 sources, it classified 16 as BL Lac objects and 3FHL J0528.4+3851 as an FSRQ. The DT classifier only provides binary probabilities, either 100\% or 0\%, i.e., a source is identified as a BL Lac if the probability is 100\% and an FSRQ if the probability is 0\%. The Random Forest classifier yielded similar results. With an accuracy of 93\%, it found 15 sources to be BL Lac objects, which is consistent with the plots in Figures \ref{fig:x+g} and \ref{fig:bla_wise}, and matches the results from the DT algorithm. The remaining two sources remain undetermined via RF method (probabilities 64\% for 3FHL J0528.4$+$3851 and 65\% for 3FHL J0648.3$+$1744) whereas DT classifies these as an FSRQ and BL Lac, respectively. \\
\indent When using the Random Forest classifier, the receiver operating characteristic curve (ROC) is used to assess the accuracy of this binary classifier. The ROC plots the true positive rate (TPR, number of true positive results) against the false positive rate (FPR, number of incorrect positive results) at differing thresholds. The accuracy is determined by finding the area under the ROC curve with 1 being the maximum value signifying all correct results. The ROC curve had an area of $\sim$0.97, signifying an accurate classifier, and is displayed below in Figure \ref{fig:roc}. \\
\indent We note that other methods exist, such as Neural Networks implemented in \cite{Chiaro2019} and \cite{Kovacevi2019}, however, we are unable to fully compare the results as only one source (3FHL J2321.6$-$1618) is shared between this work and theirs (on which they agree). 

\begin{deluxetable}{c c c c }
    \centering
    \tablecaption{Machine Learning Results \label{tab:mach} \\
    Listed are the results from the Decision Tree and Random Forest algorithms with the latter's associated probability of accurate classification. ``...'' signifies the class could not be determined at a 90\% confidence level.}
    \tablehead{\colhead{3FHL} &  \colhead{DT Pred} & \colhead{RF Pred} & \colhead{RF Prob}} 
    \startdata
J0057.9$+$6325 & bll & bll & 1.0 \\
J0233.5$+$0657$^a$ & bll & bll & 0.99 \\
J0233.5$+$0657$^b$ & bll & bll & 1.0 \\
J0359.4$-$0235 & bll & bll & 1.0 \\
J0528.4$+$3851 & fsrq & ... & ... \\
J0648.3$+$1744 & bll & ... & ... \\
J0933.5$-$5240 & bll & bll & 0.95 \\
J1439.9$-$3955& bll & bll & 1.0 \\
J1451.8$-$4145 & bll & bll & 0.96 \\
J1917.9$+$0331 & bll & bll & 1.0 \\
J1927.5$+$0153 & bll & bll & 1.0 \\
J2030.2$-$5037 & bll & bll & 1.0 \\
J2104.5$+$2117 & bll & bll & 1.0 \\
J2105.9$+$7508 & bll & bll & 0.96 \\
J2159.6$-$4619 & bll & bll & 1.0 \\
J2239.5$-$2439 & bll & bll & 1.0 \\
J2321.6$-$1618 & bll & bll & 1.0 \\
\enddata
\vspace{0.5cm}
$^a$ Left source. \\
$^b$ Right source.
\end{deluxetable}

\begin{figure}
    \centering
 \includegraphics[scale=0.4]{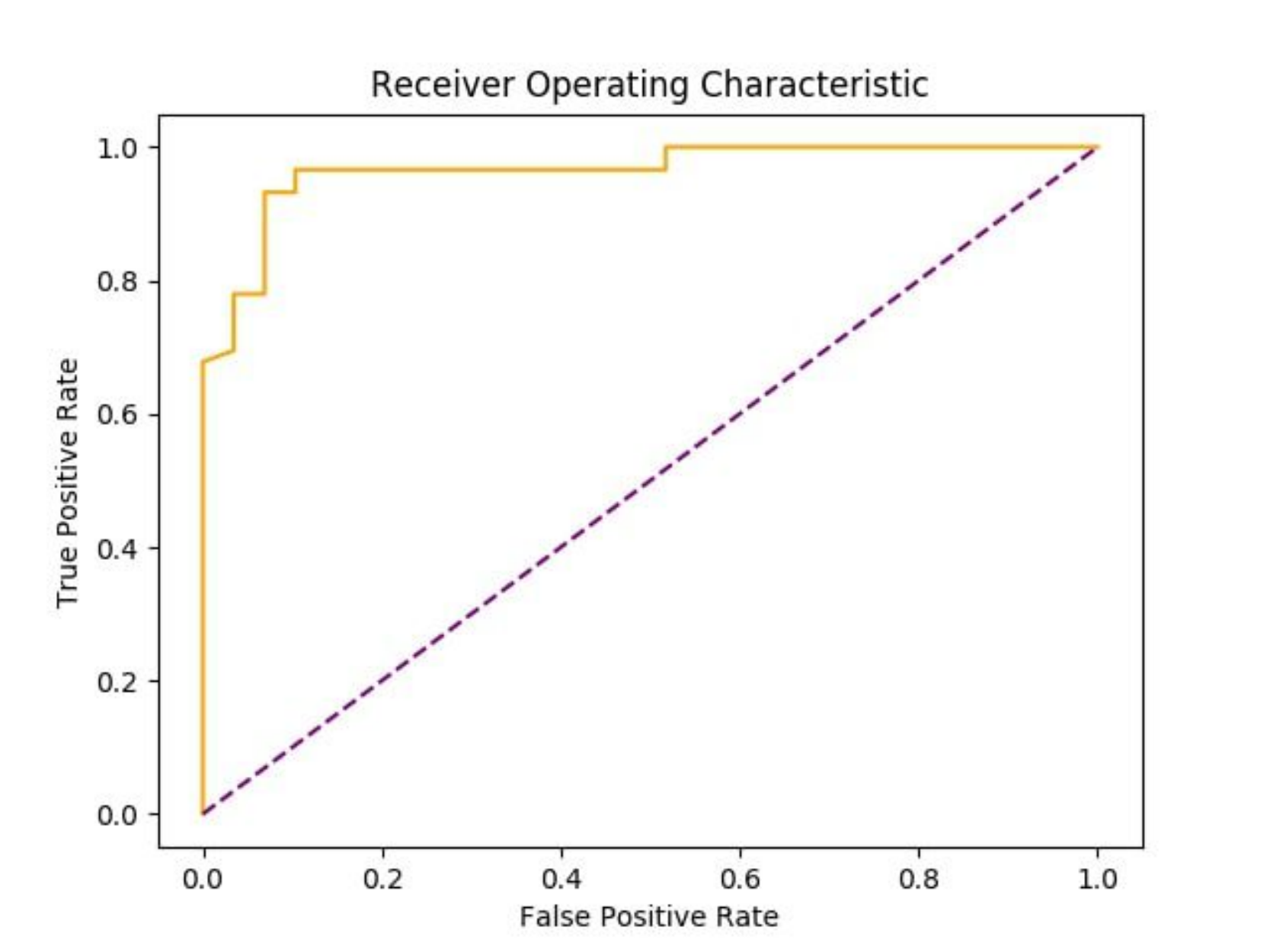}
    \caption{The ROC curve from the Random Forest method for the test sample. This curve yielded an area under the curve of 0.97. The diagonal line represents the nondiscriminatory curve, i.e., any data points on/below this line would represent non-diagnostic results.}
    \label{fig:roc}
\end{figure}

\section{Discussion and Conclusion} \label{sec:conc}
\indent \indent The primary objective of this paper was to continue the identification of all remaining 200 unclassified sources in the \textit{Fermi} 3FHL catalog. Classifying every source in the 3FHL catalog is necessary to completely understand the energetics and emission mechanisms of the high-energy universe. In this work, we analyzed \textit{Swift}-XRT observations of 38 3FHL sources and found at least one potential X-ray counterpart for 22 of them. To begin classifying our sample as Galactic or extragalactic, we compared their multiwavelength data against classified blazars and Galactic sources. While the X-ray data does not reveal any strong trend, $\gamma$-rays make it clear that our sample aligns more with blazars, in that both have a harder photon index than Galactic sources. Furthermore, the unassociated sources populate the blazar, and BL Lacs more specifically, region of the WISE blazar strip while Galactic sources have a much lower w1$-$w2 value. Due to these results, we believe the majority of these unassociated sources ($\sim19/22$) to be blazars. The remaining three sources have an uncertain nature, but are likely a star forming galaxy (3FHL J0737.5$+$6534) and pulsars (3FHL J1405.1$−$6118 and 3FHL J1907.0$+$0713).\\
\indent Towards the goal of fully classifying these sources, we implemented our machine learning algorithm to determine whether the 19 sources are BL Lacs or FSRQs. We classified these sources using their X-ray and $\gamma$-ray photon indices, WISE colors w1$-$w2 and w2$-$w3, and Variability Bayesian Blocks from the 3FHL catalog. The numerical description of how useful these parameters were, i.e., the feature importance, indicate w1$-$w2 had the greatest impact on the classification. Of the 17 sources with the necessary data, 15 were classified as BL Lacs by our Random Forest classifier and 2 (3FHL J0528.4$+$3851 and 3FHL J0648.3$+$1744) are undetermined. Using the UVOT data reported in Table \ref{tab:uvot}, we will plan observations with the SARA telescopes that will allow us to obtain redshifts and confirm the results of our classifier, thus getting one step closer to completing the 3FHL catalog. \\
\acknowledgements
The authors thank the anonymous referee for their detailed and helpful comments which greatly improved the paper. We acknowledge NASA funding under grant 80NSSC19K0151.

\bibliographystyle{aasjournal}
\bibliography{sample63}

\end{document}